\documentclass[conference]{IEEEtran}
\usepackage[utf8]{inputenc}
\IEEEoverridecommandlockouts
\usepackage{balance}
\usepackage{diagbox}
\usepackage{stackengine}
\def\delequal{\mathrel{\ensurestackMath{\stackon[1pt]{=}{\scriptstyle\Delta}}}}
\usepackage{cite}
\usepackage{mathtools} 
\DeclarePairedDelimiter\norm\lVert\rVert
\usepackage{amsmath,amsthm,amssymb,amsfonts}

\usepackage{graphicx}

\usepackage{gensymb}

\usepackage{textcomp}

\usepackage{xcolor}
\def\BibTeX{{\rm B\kern-.05em{\sc i\kern-.025em b}\kern-.08em
		T\kern-.1667em\lower.7ex\hbox{E}\kern-.125emX}}
\usepackage{epsfig}
\usepackage{mathrsfs}
\usepackage[utf8]{inputenc}
\usepackage[T1]{fontenc}
\usepackage{newtxtext,newtxmath}
\usepackage{pdfrender}
\usepackage{subfigure}
\usepackage{caption}
\usepackage{csquotes}

\def\delequal{\stackrel{\triangle}{=}}
\usepackage{mathrsfs}
\usepackage{euscript}
\usepackage{graphicx}
\usepackage{multirow}
\usepackage{algorithmicx}
\usepackage{algpseudocode}

\usepackage{cite}
\usepackage{url}
\usepackage[export]{adjustbox}
\usepackage{algorithm}
\begin{document}
	
	\title{On the Performance of Unmanned Aerial Vehicles with MIMO VLC \vspace{0.0cm}}
	\author{\IEEEauthorblockN 
		{Hosein~Zarini$^{\dag}$, Amir Mohammadisarab$^{\star}$, Maryam Farajzadeh Dehkordi$^{\star\star}$, \\Mohammad Robat Mili$^{\S}$, Bardia Safaei$^{\dag}$, Ali Movaghar$^{\dag}$, Sinem Coleri$^{\dag\dag}$ and Eduard Jorswieck$^{\S\S}$}\\$^{\dag}$Sharif University of Technology, Tehran, Iran\\$^{\star}$Universidad Miguel Hernandez de Elche, Alicante, Spain\\$^{\star\star}$George Mason University, VA, United States of America\\$^{\S}$Pasargad Institute for Advanced Innovative Solutions (PIAIS), Tehran, Iran\\$^{\dag\dag}$ Koc university, Istanbul, Turkey\\$^{\S\S}$ Technische Universitat Braunschweig, Braunschweig, Germany
	}
	\maketitle\vspace{-0.20cm}
	\begin{abstract} 
	This paper centers around a multiple-input-multiple-output (MIMO) visible light communication (VLC) system, where an unmanned aerial vehicle (UAV) benefits from a light emitting diode (LED) array to serve photo-diode (PD)-equipped users for illumination and communication simultaneously. Concerning the battery limitation of the UAV and considerable energy consumption of the LED array, a hybrid dimming control scheme is devised at the UAV that effectively controls the number of glared LEDs and thereby mitigates the overall energy consumption. To assess the performance of this system, a radio resource allocation problem is accordingly formulated for jointly optimizing the motion trajectory, transmit beamforming and LED selection at the UAV, assuming that channel state information (CSI) is partially available. By reformulating the optimization problem in Markov decision process (MDP) form, we propose a soft actor-critic (SAC) mechanism that captures the dynamics of the problem and optimizes its parameters. Additionally, regarding the high mobility of the UAV and thus remarkable rearrangement of the system, we enhance the trained SAC model by integrating a meta-learning strategy that enables more adaptation to system variations. By defining energy efficiency as a trade-off between the data rate and power consumption, simulations verify that upgrading a single-LED UAV by an array of 10 LEDs, exhibits 47\% and 34\% improvements in data rate and energy efficiency, albeit at the expense of 8\% more power consumption.
	\end{abstract}
	\begin{IEEEkeywords} 
 Multiple-input-multiple-output (MIMO), visible light communication (VLC), unmanned aerial vehicle (UAV), resource allocation, channel state information (CSI), soft actor-critic (SAC), meta-learning.
 \vspace{-0.20cm}
	\end{IEEEkeywords}
	\IEEEpeerreviewmaketitle
	\section{Introduction}
 Visible light communication (VLC) is anticipated to address the limitations of existing radio-frequency (RF) communication systems such as higher bandwidth. In VLC transmissions, data transmission is achieved using light-emitting diodes (LED) at the transmitter, while signal reception and decoding are performed by photo-diodes (PD) or cameras at the receiver \cite{VLC}. A key advantage of VLC lies in its ability to integrate illumination and communication within the same system, making it highly efficient. The deployment of multiple LEDs in a LED array has further introduced the concept of multiple-input-multiple-output (MIMO) in VLC systems, leading to significant improvements in data rates and network coverage \cite{MIMO1,MIMO2,MIMO3,sinem}. Later, the application of VLC band in aerial communication platforms, particularly through the use of unmanned aerial vehicles (UAVs) equipped a single-LED. This achievement promised even better coverage and flexible maneuverability over terrestrial VLC systems \cite{Hosein}. To date, a plethora of literature have studied various aspects of VLC-enabled UAV-based networks, including network coverage \cite{cov}, data rate \cite{data_rate} and energy efficiency \cite{EE} standpoints. Nonetheless, despite the tremendous gain brought by MIMO-VLC \cite{MIMO1,MIMO2,MIMO3}, its application to UAVs remains largely unexplored in the existing literature \cite{Hosein,cov,data_rate,EE}. 
  \par This paper aims to address the identified research gap by proposing a resource allocation mechanism to evaluate the performance of a UAV equipped with an LED array, forming UAV-MIMO-VLC system.
    Two key design challenges are encountered in this context: $i)$ LED arrays consume more power compared to a single-LED, raising concerns due to the UAV's strict battery limitation. $ii)$ The considerable dynamics of the UAV-MIMO-VLC system originated from frequent motion of the UAV poses a challenge for real-time and on-demand resource allocation.
    %To suppress this challenge, we leverage the concept of multi-domain hybrid dimming, introduced in \cite{} and concurrently optimize the quantity of glaring LEDs in spatial dimming (SD) and the direct current (DC) bias level in analog dimming (AD) to achieve a specified dimming level. This paper unlike prior studies based on convex optimization tools\cite{} and deep reinforcement learning (DRL) tools \cite{}, proposes a real-time resource allocation based on meta-learning that effectively captures the system variations. 
  Toward addressing these challenges, this paper proposes the following contributions
  \begin{itemize}
      \item This is the first endeavor studying a UAV equipped with a LED array. Regarding the battery limitations of the UAV, we adopt a hybrid dimming control strategy on the LED array that jointly manages the direct current (DC)-bias level and the number of glaring LEDs at the LED array to substantially reduce its energy consumption. 
      \item We present a real-time and adaptive, yet robust resource allocation scheme to evaluate this system. Specifically, under the assumption of imperfect channel state information (CSI), a soft actor-critic (SAC) agent is trained to optimize the system performance. Due to frequent mobility of the UAV and thus considerable dynamism of the system, we fine-tune the training process of the SAC agent with meta-learning to enhance its adaptability and generalization.
      \item Numerical results demonstrate that the proposed UAV equipped with a 10 LEDs considerably outperforms the conventional single-LED UAVs studied in \cite{Hosein,cov,data_rate,EE}. Specifically, by defining energy efficiency as a trade-off between data rate and power consumption, our proposed system achieves a 47\% and 34\% advancement in data rate and energy efficiency, respectively. Nevertheless, this progress comes with a marginal trade-off, as the system incurs an 8\% increase in power consumption.
  \end{itemize}
  
%In what follows, we organize the paper by introducing the UAV-MIMO-VLC system configuration and its corresponding optimization problem in Section \ref{sysModel}. We continue the paper by elaborating our solution approach based on SAC integrated with meta-learning in Section \ref{Solution}. Simulation results in Section \ref{Simulation} verify the effectiveness of the proposed system and the proposed resource allocation strategy. The paper is eventually concluded in Section \ref{Conclusions} by drawing insights for future research directions.
\begin{figure} 	
 	\centering
 	\includegraphics[scale=0.26]{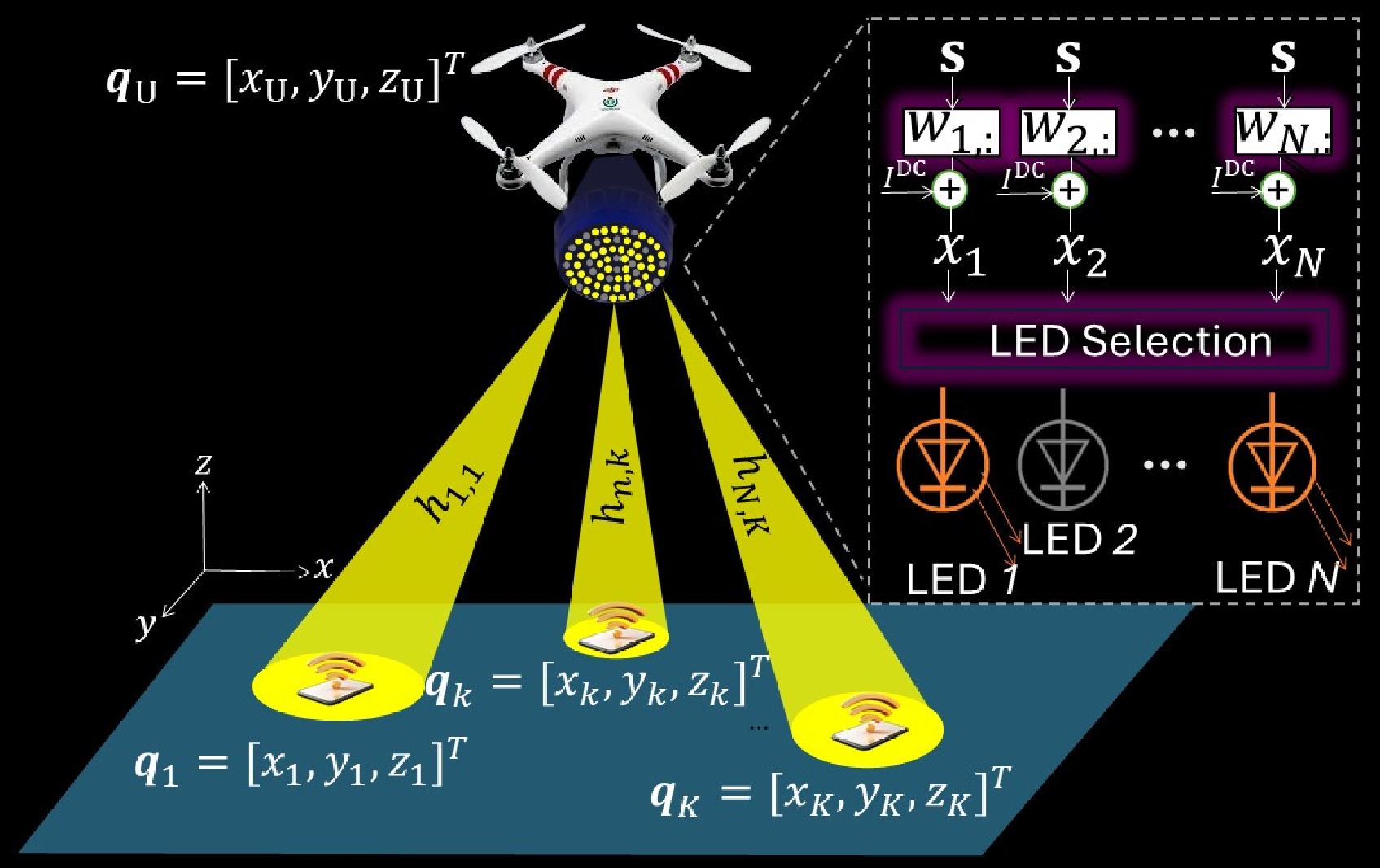}
 	\caption{The considered UAV-MIMO-VLC system.}\vspace{-0.5cm}
 	\label{fig:SysModel} 
\end{figure}
\section{System Model and Problem Formulation} 
\label{sysModel}
In this section, we elaborate the UAV-MIMO-VLC system setting and formulate an optimization problem for evaluating the performance of this system.
\subsection{Signal Model}
Downlink transmission of a visible light communication (VLC) system is studied as depicted in Fig.~\ref{fig:SysModel}, within which an array of LEDs is patched to a rotary-wing unmanned aerial vehicle (UAV) for the illumination and communication purpose at the same time. In specific, the LED array includes a set $\mathcal{N} = \left\{ {1,2,...,{N}} \right\}$ of ${N}$ LEDs deployed at the UAV to serve a set $\mathcal{K} = \left\{ {1,2,...,{K}} \right\}$ of $K$ randomly positioned users, each equipped with a single photo-detector (PD). The LEDs are supposed to be independently modulated via separate drivers, yet all are deployed at the UAV as a central controller that collects channel feedback and performs resource management.  Let ${\textbf{s}}=\left[s_1,s_2,...,s_K\right]\in \mathbb{R}^{K\times1}$ denote the data symbol vector, intended for all users, such that $s_k\in\left[-1,1\right]$ specifies the data symbol for the user $k$ and $\mathbb{E}\{|{{s}}_{k}|^2\}$~=~1 $\forall k\in \mathcal{K}$. Also, let us define a transmit precoding matrix $\textbf{W}=\left[\textbf{w}_1,\textbf{w}_2,...,\textbf{w}_K\right]\in\mathbb{R}^{N\times K}$ with $\textbf{w}_k=\left[w_{1,k}, w_{2,k}, ..., w_{N,k}\right]\in\mathbb{R}^{N\times1}$ being the beamforming vector for the user $k$. Designing an efficient precoding scheme significantly helps controlling the inter-user interference. Without loss of generality, we assume a linear precoding for data symbols\cite{Hosein}. In optical communications, generated signals have to be real-valued and non-negative. This requirement is covered by direct current (DC)-biased modulation, through which the brightness of LEDs and average optical power are also determined. Taking this fact into consideration, the transmit signal vector of all LEDs can expressed as $\bold{x}= \sum_{k=1}^{K}{\bold{w}}_k{\textbf{s}}_k+\textbf{I}^{\text{DC}},$ where the DC bias ${\textbf{I}^{\text{DC}}} = \left[I^{\text{DC}}_{1}, I^{\text{DC}}_{2}, ..., I^{\text{DC}}_{N}\right]\in\mathbb{R}^{N\times 1} $ has been incorporated so as to ensure that the amplitude of the transmit signal vector resides within the non-negative range of LEDs. Besides, for the sake of uniformity of illumination in indoor environments, the DC bias is assumed to be the same for all the LEDs\cite{HD-1}. On this basis, the LED indices can be simply removed, which results in $I^{\text{DC}}_{n}=I^{\text{DC}}~\forall n\in\mathcal{N}$ and ${\textbf{I}^{\text{DC}}} = I^{\text{DC}}$. Accordingly, the transmit signal of a typical LED $n$, i.e., $x_n$ satisfies $-\sum_{k=1}^{K}|w_{n,k}|+I^{\textrm{DC}}\le x_{n}\le\sum_{k=1}^{K}|w_{n,k}|+I^{\textrm{DC}}~\forall n\in\mathcal{N}.$ All LEDs, however, work within their dynamic ranges to avoid signal clipping. This consideration poses a constraint for the transmit beamforming vector as $\sum\limits_{k = 1}^K {\left| {{\bold{w}_k}} \right|}  \le \min ({I^{\textrm{DC}}} - {I_l},{I_h} - {I^{\textrm{DC}}})~\forall n\in \mathcal{N}$, with ${I_l}$ and ${I_h}$ being the lower and upper bound of the LED driver current, respectively. 
\subsection{Channel Model}
For a typical VLC receiver, a line-of-sight (LoS) link constitutes about 95\% of the total received power, such that at least 7 dB difference is roughly observed between the received power for the LoS link and that for the most dominant non-line-of-sight (NLoS) one\cite{Hosein}. Given this significant disparity, we simplify our analysis by focusing exclusively on the LoS links and disregarding the NLoS components. Suppose that the UAV and the typical user $k$ are located at the coordinates of $\bold{q}_{\textrm{U}}=[{x}_{\textrm{U}},{y}_{\textrm{U}},{z}_{\textrm{U}}]^T$ and $\bold{q}_k=[{x}_k,{y}_k, {z}_k]^T$, respectively. In comparison with the distance between the UAV and ground users, the distance between the LEDs deployed at the UAV is negligible. Hence, a unified (and not per LED) coordinate is considered for the UAV. So, the direct distance between the user $k$ and the UAV can be calculated as: $d_{\textrm{U},k}=\sqrt{(x_{\textrm{U}}-{x}_{k})^{2}+(y_{\textrm{U}}-{y}_{k})^{2}+(z_{\textrm{U}}-{z}_{k})^{2}}$. However, the channel condition can vary significantly due to differences in their transmission semi-angles. We denote the optical channel gain from the LED $n$ to the user $k$ by $h_{n,k}\in \mathbb{R}$, which is modelled as: $	h_{n,k}\!=\frac{(m+1)A_k}{2\pi d_{\textrm{U},k}^{2}}G^{\textrm{VLC}}(\psi_{n,k})\!\cos^{m}(\phi_{n,k})\!\cos(\psi_{n,k})$ for $0\!\leq\!\psi_{n,k}\!\leq\!\Psi_{c}$, and $h_{n,k}=0,$ otherwise. In such a channel model, the detection area of the PD at the user $k$ is defined by $A_k$. Besides, $m=-\frac{\ln2}{\ln(\cos\Phi_{1/2})}$ represents the order of Lambertian emission where $\Phi_{1/2}$ denotes the half-power semi-angle of the LED. The angles of incidence and irradiance between the LED $n$ and the user $k$, are respectively represented by $\psi_{n,k}$ and $\phi_{n,k}$, such that $\cos (\psi_{n,k}) = \cos (\phi_{n,k}) = d_{\textrm{U},k}/(z_{\textrm{U}}-{z}_{k})$. Moreover, $\Psi_{k}$ specifies the field of vision (FOV) semi-angle at the user $k$ and $G^{\textrm{VLC}}(\psi_{n,k})$ indicates the gain of the optical concentrator, %defined as $G^{\textrm{VLC}}(\psi_{n,k})=\frac{q^{2}}{\sin^{2}\left(  \Psi_{n,k}\right)}$ for $0\leq\psi_{n,k}\leq\Psi_{k}$ and $G^{\textrm{VLC}}(\psi_{n,k})=0$, otherwise,
%    	\begin{equation}
%    	G^{\textrm{VLC}}(\psi_{n,k})=\left\{
%    	\begin{array}
 %   	[c]{c}%
%    	\frac{q^{2}}{\sin^{2}\left(  \Psi_{n,k}\right)  }\\
%    	0,
%    	\end{array}
%    	\right.  \left.
%    	\begin{array}
%    	[c]{c}%
%    	0\leq\psi_{n,k}\leq\Psi_{k},\\
%    	\psi_{n,k}>\Psi_{k},%
%    	\end{array}
%    	\right.
%    	\end{equation}
    	with $q\ge0$ introducing the internal refractive index. 
     However, the availability of perfect CSI is an ideal assumption, which leads to remarkable performance degradation. Especially in VLC systems, where optical lightweight signals are heavily prone to environmental obstacles or reflections, the practical consideration of imperfect CSI at the transmitter (the UAV in our scenario) provides the system with a realistic performance analysis \cite{Imperfect}. We model the imperfect channel between the LED $n$ and user $k$ as $\hat{h}_{n,k} = h_{n,k} + \epsilon_{n,k},$ wherein $\epsilon_{n,k}$ represents the estimation error within a bounded region. This error is defined as $|\epsilon_{n,k}| \leq \delta$, where $\delta$ corresponds to the channel uncertainty radius and is assumed be a small constant.
\subsection{Hybrid Dimming}
As discussed earlier, the concept of hybrid dimming is a multi-domain dimming approach that exploits analog domain (AD) and spatial domain (SD) at the same time\cite{HD-1}. By utilizing SD, only a subset of LEDs is glared, while the others remain inactive, resulting in significantly reduced power consumption compared to traditional systems that rely solely on digital dimming (DD). For equipping UAVs with optical multiple-input-multiple-output (MIMO) technology using a LED array, implementing hybrid dimming control scheme is essential, given the strict energy limitation of the UAV. To this purpose, let us introduce a binary LED selection matrix $\mathbf{A} = \text{diag}(\mathbf{a}) \in \lbrace 0, 1 \rbrace^{N \times N},$ with $\mathbf{a} = [a_{1}, \dots, a_{N}]^T \in \lbrace 0, 1 \rbrace^{N \times 1}$, such that $a_n$=1 specifies the glared state of the LED $n$, whereas $a_n$=0 delineates that it is inactive. Accordingly, the number of active LEDs can be calculated as ${N_{\textrm{a}}} = \sum\limits_{n = 1}^{{N}} {{a}_n}$.
By adopting a hybrid dimming control mechanism, the signal benefits from both analog and spatial domains simultaneously. So, the uniform DC-bias level (corresponded to AD) and the number of glared LEDs (corresponded to SD) need to be jointly adjusted. To do so, we firstly define a target dimming level denoted by $\eta$ with a predetermined value. Given $\eta$ then, the number of glared (actual working) LEDs for SD can be simply rounded-off as $N_a=\big[\eta N\big].$ On this basis, the uniform DC-bias level $I^{\textrm{DC}}$ for AD will be acquired as\cite{HD-1}: $I^{\textrm{DC}}=\dfrac{\eta N(I_0-I_l)}{N_a}+I_l$, in which $I_0=\dfrac{I_l+I_h}{2}$ defines the original DC-bias for AD, when all $N$ LEDs are glared and $\eta= 100\%$ dimming level is set. %Reminded that DC optical orthogonal frequency division multiplexing (DCO-OFDM) achieves its best performance, when the DC-bias resides at the middle of the dynamic range\cite{HD-1}. 
By determining ($N_{\textrm{a}}$) thus far, we are aware of how many LEDs are glared in fact. However, it is required to exactly specify which LEDs at the LED array have to be active. This will be clarified by optimizing the binary LED selection matrix $\textbf{A}$ in a network-wide optimization problem, which will be mathematically formulated in upcoming subsections.
\subsection{Trajectory Control and Power Consumption for UAV}
For a durable and efficient aerial communication, it is required to meticulously trace the flight trajectory and power consumption of UAV. 
\par \textbf{Trajectory Control:} Concerning the high mobility of the UAV and for the sake of analysis, we discretize its overall flight time duration $T$ to a set $\mathcal{L}=\left\{ {0,1,2,...,L} \right\}$ with $L$ time slots, each occupying a duration of ${\tau}$, such that $T=L\tau$. The duration of time slots is chosen to be sufficiently small, ensuring that the configuration of the network remains static and the optical channels follow a quasi-static model.
Within a typical time slot $l$, the UAV is assumed to fly with the velocity of ${\bold{v}_{\textrm{U}}}\left( l \right) = {\left[ {v_{\textrm{U}}^x\left( l \right),v_{\textrm{U}}^y\left( l \right),v_{\textrm{U}}^z\left( l \right)} \right]^T}\in \mathbb{R}^{3 \times 1}$, such that its initial and final positions are the same, i.e., it returns to the beginning position at the end of the last time slot $L$ for recharging.
The flight restrictions of the UAV impose several constraints on the system. First, ${\bold{q}_{\textrm{U}}}\left( {l + 1} \right) = {\bold{q}_{\textrm{U}}}\left( l \right) + {\bold{v}_{\textrm{U}}}\left( l \right){\tau}$ determines its next location with respect to (w.r.t.) the movement velocity.
Through ${\bold{q}_{\textrm{U}}}\left[ 0 \right] = {\bold{q}_{\textrm{U}}}\left[ L \right]={\bold{q}_{\textrm{U}}^{\textrm{I}}},$ we guarantee the return of the UAV to its initial location ${\bold{q}_{\textrm{U}}^{\textrm{I}}}$ at the end of the last time slot $L$. We restrict the UAV flight to the minimum and maximum boundaries ${\bold{q}_{\min }}$ and ${\bold{q}_{\max }}$ as
${\bold{q}_{\min }} < {\bold{q}_{\textrm{U}}\left( l \right)} < {\bold{q}_{\max }}$. Finally,
$\left\| {{\bold{v}_{\textrm{U}}}\left( {l + 1} \right) - {\bold{v}_{\textrm{U}}}\left( l \right)} \right\| \le a _{\text{max}}{\tau}$ and $\left\| {{\bold{v}_{\textrm{U}}}\left( l \right)} \right\| \le V_{\text{max}}$ control the flight acceleration and velocity of the UAV by introducing the maximum flight acceleration $a _{\text{max}}$ and maximum flight velocity $V _{\text{max}}$, respectively. 
\par \textbf{Power Consumption:} The power consumed during the hovering of the UAV can be categorized into blade power consumption $P_b$ and induced power consumption $P_i$, modelled as~\cite{Rui}: ${P_{\textrm{Hov}}} = P_b+P_i$ where $P_b \delequal\frac{\rho }{8}\zeta \delta {A_{\textrm{U}}} \Omega_{\textrm{U}}^{3} R_{\textrm{U}}^3$ and ${P_i}\delequal\left( {1 + \iota } \right)\frac{{W_{\textrm{U}}^{3/2}}}{{\sqrt {2\zeta {A_{\textrm{U}}}} }}$. In this context, $P_{\text{Hov}}$ includes the profile drag coefficient $\rho$, the air density $\zeta$, the rotor solidity $\delta$, the blade angular velocity $\Omega_{\textrm{U}}$ in radian per second, the rotor disk area $A_{\textrm{U}}$ in square meters ($\textrm{m}^{2}$), the rotor radius $R_{\textrm{U}}$ in meter, the incremental correction factor $\iota$ and the UAV weight $W_{\textrm{U}}$ in Newton. Accordingly, the propulsion power consumption of the UAV, encompassing  the blade profile, the induced, and the parasite power is given by:
${{\mathop{\rm P}\nolimits} _{\text{Prop}}} =\underbrace {\left( {1 + \frac{{3 \norm{\textbf{v}_{\textrm{U}}}^{2}    }}{  \Omega_{\textrm{U}}^{2}  R_{\textrm{U}}^{2}    }} \right){P_b}}_{\text{blade profile}} + \underbrace {\left( {\sqrt {\sqrt {1 + \frac{{\norm{\textbf{v}_{\textrm{U}}}^{4} }}{{4v_{\textrm{UI}}^4}}}  - \frac{{\norm{\textbf{v}_{\textrm{U}}}^{2} }}{{2v_{\textrm{UI}}^2}}} } \right){P_i}}_{\text{induced}}\\+ \underbrace {\frac{1}{2}{d_\text{U}}{\zeta \delta }{A_\text{U}}\norm{\textbf{v}_{\textrm{U}}}^{3} }_{\text{parasite}},
$ in which $v_{\textrm{UI}}$ specifies the mean rotor induced velocity in hover, while $d_{\textrm{U}}$ denotes the fuselage drag ratio.
Finally, the total consumed power of the system can be calculated as: $P^{\text{Tot}}=\ddot{\zeta}\sum_{n=1}^{N}\sum_{k=1}^{K}\mathbf{a}{w}_{n,k}+\varphi {N_a}{I^{\textrm{DC}}}+P^{\text{Cir}}+P^{\text{Prop}},$ with $\ddot{\zeta}$ being the amplifier efficiency factor, $\varphi$ determining the conversion factor, as well as $P^{\text{Cir}}$ designating the consumed power of switch and control circuit modules at the UAV.
\subsection{Transmission Protocol}
%By breaking the orthogonality of resources and adopting a non-orthogonal approach for resource allocation, NOMA achieves numerous benefits for VLC systems, some of them such as higher data rate, massive connectivity, better utilization of resources, lower bit error rate (BER), have been previously highlighted in literature\cite{Opp}. However, the prior HD-enabled VLC systems\cite{HD-1,HD-2,HD-3,HD-4} have never gone beyond the orthogonality of resources.
%By superimposing multiple users over the same resource on the basis of NOMA, users will be served via different power levels according to their channel conditions. The PD-equipped users as the receiver, next employ successive interference cancellation (SIC) and decode their intended signal effectively. 
We adopt the power-domain non-orthogonal multiple access (NOMA) scheme for enabling massive access in the system. Suppose $	{{\left|\bold{h}^{H}_{1}{\bold{w}}_{1} \right|}^2}\le{{\left|\bold{h}^{H}_{2}{\bold{w}}_{2} \right|}^2}\le...\le {{\left|\bold{h}^{H}_{K}{\bold{w}}_{K} \right|}^2}$ as the descending order of the channel coefficients. 
    	%    	\begin{align}
    	%    	{\left| {{h_1}} \right|^2} \le {\left| {{h_2}} \right|^2} \le ... \le {\left| {{h_K}} \right|^2}
    	%    	\end{align}
    	Then, the instantaneous received data rate of the user $k$ can be expressed as:
    	\begin{align} \label{rate}
    		R_{k}(\mathbf{W},\mathbf{A},\mathbf{Q})= \log_{2}\bigg(1+\dfrac{|\bold{h}^{H}_{k}\bold{A}{\bold{w}}_{k}|^{2}}{\sum_{i=k+1}^{K}|\bold{h}^{H}_{k}\bold{A}{\bold{w}}_{i}|^{2}+\sigma^{2}_{k}}\bigg),
    	\end{align}
    	where ${\bold{h}_k} = {\left[ {{h_{k,1}},{h_{k,2}},...,{h_{k,{N}}}} \right]^T}$ denotes the channel vector of the user $k$. Accordingly, we can define energy efficiency as a trade-off between the aggregate data rate of the network and its overall power consumption as follows.
     \begin{align}\label{EE}
         EE(\mathbf{W},\mathbf{A},\mathbf{Q})= \dfrac{\sum_{k}R_{k}(\mathbf{W},\mathbf{A},\mathbf{Q})}{P^{\textrm{Tot}}}.
     \end{align}
\subsection{Problem Formulation}
%Regarding the strict energy limitation of the UAV on one hand and notable power consumption of the LED array on the other hand, 
We aim to jointly optimize the network resources $\big\{\mathbf{W},\mathbf{Q} = \{ {\bold{q}_{\textrm{U}}}, {{\bold{v}_{\textrm{U}}}}\},\bold{A}\big\}$, such that the power consumption of the UAV is minimized and the UAV-MIMO-VLC system requirements are met. Mathematically, this problem is formulated as follows:
    	\begin{subequations}
    		\label{QoE_max_prob.1}
    		\begin{align}
    		\mathcal{P}_{1}:\nonumber&\min_{\mathbf{W},\mathbf{A},\bold{Q}}{P^{\text{Tot}}}
    		\\&\text{s.t.}~~~~\!\text{C}_{1}:{R_{k}}(\mathbf{w},\mathbf{A},\mathbf{Q})\ge {R_{\text{min}}},~\forall k\in \mathcal{K},
    		\\&~~~~~~~\text{C}_{2}: P^\text{Tot}\leq P_{\textrm{max}},
    		\\&~~~~~~~ \text{C}_{3}:\sum\limits_{k = 1}^K {\left| {{\bold{{w}}_k}} \right|}  \le \min ({I^{\textrm{DC}}} - {I_l},{I_h} - {I^{\textrm{DC}}}), \forall n \in \mathcal{N},\\
    			&~~~~~~~ \text{C}_{4}:{\bold{q}_{\textrm{U}}}\left( {l + 1} \right) = {\bold{q}_{\textrm{U}}}\left( l \right) + {\bold{v}_{\textrm{U}}}\left( l \right){\tau}, 
    			\quad \forall l\in{\mathcal{L}}, \\%    			&~~~~~~~{{\left|\bold{h}^{H}_{1}{\bold{w}}_{1}\right|}^2}\le...\le {{\left|\bold{h}^{H}_{K}{\bold{w}}_{K} \right|}^2},\\
    			&~~~~~~~\text{C}_{5}:{\bold{q}_{\min }} < {\bold{q}_{\textrm{U}}\left( l \right)} < {\bold{q}_{\max }}, \quad \quad \forall l\in{\mathcal{L}}, \\
    			&~~~~~~~\text{C}_{6}:\left\| {{\bold{v}_{\textrm{U}}}\left( {l + 1} \right) - {\bold{v}_{\textrm{U}}}\left( l \right)} \right\| \le a _{\text{max}}{\tau}, \quad \forall l\in{\mathcal{L}},  \\
    			&~~~~~~~\text{C}_{7}:\left\| {{\bold{v}_{\textrm{U}}}\left( l \right)} \right\| \le V_{\text{max}}, \quad \quad \forall l\in{\mathcal{L}},  \\
    			&~~~~~~~\text{C}_{8}: \eta=\dfrac{{N_{\textrm{a}}}(I^{\textrm{DC}}-I_l)}{N(I_0-I_l)}\times100\%,\\
    		    &~~~~~~~\text{C}_{9}: a_{n} \in{\{0,1\}},~~\forall n \in \mathcal{N}.
    		\end{align}
    	\end{subequations}
    	The quality-of-service (QoS) requirement of all users ($R_{\textrm{min}}$) is ensured in $\text{C}_{\text{1}}$. $\text{C}_{\text{2}}$ guarantees the transmit power budget of the UAV denoted by ($P^{\textrm{max}}$). In $\text{C}_{\text{3}}$, the dynamic range of all LEDs in the LED array is bounded. The flight restrictions of the UAV are described in $\text{C}_{\text{4}}$-$\text{C}_{\text{7}}$ as discussed earlier. The target dimming level $\eta$ is assured to be served in $\text{C}_{\text{8}}$. $\text{C}_{\text{9}}$ finally defines the binary domain of the LED selection variable. The coupling of variables from both continuous and discrete domains along with the inclusion of the non-convex constraint $C_{\text{1}}$ renders this problem non-convex. This observation categorizes $\mathcal{P}_1$ as a mixed-integer and nonlinear programming (MINLP), thereby classifying it as non-deterministic polynomial-time hardness (NP-hard). Unlike classical convex optimization-based solutions that offer intricate mathematical transformation to solve non-convex optimization problems, we propose a real-time and adaptive solution strategy to solve $\mathcal{P}_1$ in the following section.
     \section{Proposed Meta-SAC Based Solution Strategy} \label{Solution}
     In this section, we reformulate \eqref{QoE_max_prob.1} in MDP form and propose a Meta-SAC algorithm that integrates meta-learning with SAC. Compared to existing learning-driven resource allocation schemes mostly based on deep reinforcement learning (DRL) \cite{Hosein}, the integration of meta-learning in our proposed method allows more adaptability to system variations, thereby yielding a more real-time and flexible solution framework. 
     \subsection{Reformulation of $\mathcal{P}_1$ in MDP form}
     We consider the UAV to function the central controller, acting as a DRL agent, interacting with the considered MIMO-VLC system, which serves the DRL environment. The agent selects an action \( \textbf{a}(l)=\{\textbf{W}(l),\textbf{A}(l),\textbf{Q}(l)\} \) in time step \( l \) based on the current state \( \textbf{s}(l)=\{\textbf{h}_{k}\},~\forall k\in \mathcal{K}, \) of the environment. We view \( P_r(\textbf{s}(l+1) | \textbf{s}(l), \textbf{a}(l)) \) as the transition probability, representing the likelihood of transitioning to a new state \( \textbf{s}(l+1) \) from the current state of the environment \( \textbf{s}(l) \), when action \( \textbf{a}(l) \) is executed. Consequently, the environment transits to the new state \( \textbf{s}(l+1) \), and a reward function \( Re(\textbf{s}(l), \textbf{a}(l)) \) evaluates the efficacy of the chosen action. Formally, the reward function is defined as $Re(\textbf{s}(l), \textbf{a}(l)) = -P^{\text{Tot}}$, when $\text{C}_{\text{1}}$-$\text{C}_{\text{9}}$ are met and $Re(\textbf{s}(l), \textbf{a}(l))=0$, otherwise.
%     \begin{align}
%Re(\textbf{s}(l), \textbf{a}(l)) = \begin{cases}
%-P^{\text{Tot}} & \text{if  $\text{C}_{\text{1}}$-$\text{C}_{\text{9}}$, are satisfied,}\\
%-10&\text{otherwise}.
%\end{cases}
%\end{align}
\subsection{Proposed Meta-SAC Algorithm}
\par As a robust DRL algorithm tailored for dynamic scenarios, SAC \cite{SAC} is equipped with an agent whose objective is to determine the optimal policy $\pi^*$ that maximizes a trade-off between the expected cumulative reward over time and a certain level of uncertainty or exploration in its actions, known as entropy. This trade-off measures the policy's randomness and can be formulated as: $\pi^* = \arg\max_{\pi} \mathbb{E}_{(\mathbf{s}(l), \mathbf{a}(l)) \sim \mathcal{P}} \bigg[ \sum_{l=0}^{\infty} \gamma^l Re(\mathbf{s}(l), \mathbf{a}(l)) - \lambda \sum_{l=1}^{\infty} \gamma^l \log \left( \pi(\cdot|\mathbf{s}(l)) \right)  | \mathbf{s}(l)=\mathbf{s}(0), \mathbf{a}(l)=\mathbf{a}(0)\bigg],$
within which $\gamma \in (0, 1]$ designates the discount factor and $\mathcal{P}$ stands for the transition probabilities. While effective in dynamic settings, where the training and testing environments of the DRL agent match, the traditional SAC struggles to swiftly adapt to rapidly changing conditions, such as those encountered with a flying UAV. To tackle this situation, we introduce the Meta-SAC approach, by integrating the model-agnostic meta-learning (MAML) \cite{MAML} with SAC. This integration enhances Meta-SAC's generalization capabilities, enabling rapid adaptation to previously unseen environments \cite{TVT}. Whereas the traditional SAC employs one actor and two critic networks, characterized by parameters $\phi$, $\psi_1$, and $\psi_2$, respectively, meta-learning integrates one target actor ($\phi_e$) and two target critic networks ($\psi_{e1}$ and $\psi_{e2}$) to facilitate convergence.
%Meta-learning has proven its efficacy in multi-task scenarios by extracting task-agnostic knowledge from a collection of tasks. This acquired knowledge can subsequently be utilized to improve the learning process for new tasks within the same collection [16]. In the following sections, we'll elaborate our proposed Meta-SAC method.
\par The system model outlined in Section \ref{sysModel} incorporates random placements of the UAV and users in each instantiation. Consequently, the optimization problem $\mathcal{P}_1$ can be considered as a meta task scenario. Here, each task is defined by assuming distinct initial random locations for the UAV and users. Following the MAML approach \cite{MAML}, we define a meta task set $\mathcal{T}_t = \{t = 1, 2, \ldots , T\}$, in which each task $t \in \mathcal{T}_t$ can be represented by the tuple $(\textbf{s}_t(l), \textbf{a}_t(l), \textbf{s}_t(l+1), r_t(l)),~\forall l\in\mathcal{L}$. Moreover, each task $t$ possesses its specific replay buffer labeled as $D_t$, a support set $D_{t}^{tr}$ as well as a query set $D_{t}^{val}$. The proposed Meta-SAC approach consists of two main stages, namely Meta-training and Meta-adaptation.
\par \textbf{Meta-Training Phase:} In the initial stage of meta-learning, both the actor and critic networks undergo training using a two-tier approach that encompasses individual-level and global-level updates \cite{TVT}. At the former level, for each task $t$, the environment resets to initialize the state $\textbf{s}_t(0)$. Next, by adopting and applying the action $\textbf{a}_t(l)$ in each time step $l$, the agent updates the environment to the new state $\textbf{s}_t(l+1)$ and the reward function $Re_t(l)$ evaluates the efficacy of the chosen action. Relying on adaptive moment estimation (ADAM) algorithm, subsequently, the actor and critic networks' parameters for each task $t$, denoted by $\hat{\phi}_t$, $\hat{\psi}_{1,t}$, and $\hat{\psi}_{2,t}$ respectively, are updated as: $\hat{\phi}_t=\arg\min_{\phi} \mathcal{L}_t^A(\phi, D_t^{tr})$, $\hat{\psi}_{1,t}=\arg\min_{\psi_1} \mathcal{L}_t^{C1}(\psi_1, D_t^{tr}),$ and  $\hat{\psi}_{2,t}=\arg\min_{\psi_2} \mathcal{L}_t^{C2}(\psi_2, D_t^{tr})$,
with $\mathcal{L}_t^A(\phi, D_t^{tr})$, $\mathcal{L}_t^{C1}(\psi_1, D_t^{tr})$ and $\mathcal{L}_t^{C2}(\psi_2, D_t^{tr})$ quantifying the loss functions for the actor and critic networks, respectively. After applying individual-level updates of all tasks, the next step involves updating the global network parameters at the global-level. This process entails sampling a query set $D_t^{val}$ randomly from the dataset $D_t$. Accordingly, the global network parameters are updated as: $\phi=\arg\min_{\phi} \sum_{t \in \mathcal{T}_t} \mathcal{L}_t^A(\hat{\phi}_t, D_t^{val})$, 
$\psi_1=\arg\min_{\psi_1} \sum_{t \in \mathcal{T}_t} \mathcal{L}_t^{C1}(\hat{\psi}_{1,t}, D_t^{val})$ and $\psi_2=\arg\min_{\psi_2} \sum_{t \in \mathcal{T}_t} \mathcal{L}_t^{C2}(\hat{\psi}_{2,t}, D_t^{val})$.
\par \textbf{Meta-Adaptation Phase:}
During this stage, we insightfully initialize the parameters of the actor and critic networks relying on the trained parameters from the prior stage. In particular, we initialize $\Phi$, $\Psi_1$, and $\Psi_2$, the parameters of the actor and critic networks, as $\Phi_0 = \phi$, $\Psi_{1,0} = \psi_1$, and $\Psi_{2,0} = \psi_2$, respectively. This stage already exploits another replay buffer $D_{ada}$, from which batch of data $\mathbb{D}_{ada}$ are randomly drawn from. Accordingly, the update rules for the parameters of the actor and critic networks are respectively expressed as:
$\Phi=\Phi - \beta_A \nabla_{\Phi} \mathcal{L}_A(\Phi, D_{ada}),$
$\Psi_1=\Psi_1 - \beta_{C1} \nabla_{\Psi_1} \mathcal{L}_{C1}(\Psi_1, D_{ada}),$ and $\Psi_2=\Psi_2 - \beta_{C2} \nabla_{\Psi_2} \mathcal{L}_{C2}(\Psi_2, D_{ada})$, with $\beta_A$, $\beta_{C1}$, and $\beta_{C2}$ denote the learning rates for the actor and critic networks respectively. The computational complexity for the Meta-SAC algorithm depends on the number of layers at the SAC \cite{TVT}.
%\subsection{Computational Complexity}
%The computational demand of Meta-SAC incorporates both training and running phases. During training, this complexity is stated as $O\left(\left(\sum_{l=0}^{L} h_l h_{l+1}\right) \times D_{t}^{tr} \times N_{\text{time-step}} \times N_{\text{task}} \times N_{\text{episode}}\right).$ Similarly, the running phase poses the computational complexity from the order of
%$
%O\left(\left(\sum_{l=0}^{L} h_l h_{l+1}\right) \times D_{t}^{tr} \times N_{\text{time-step}} \times N_{\text{episode}}\right).$
%Here, we represent the number of time steps, tasks, and episodes, respectively by $N_{\text{time-step}}$, $N_{\text{task}}$, and $N_{\text{episode}}$. More so, we respectively designate the number of neural network layers, neurons in each layer, and neurons in the $l$-th layer by $L$, $h_l$, and $h_{l+1}$.
\begin{figure} 	
 	%\hspace{-0.6cm}
 	\includegraphics[scale=0.1550]{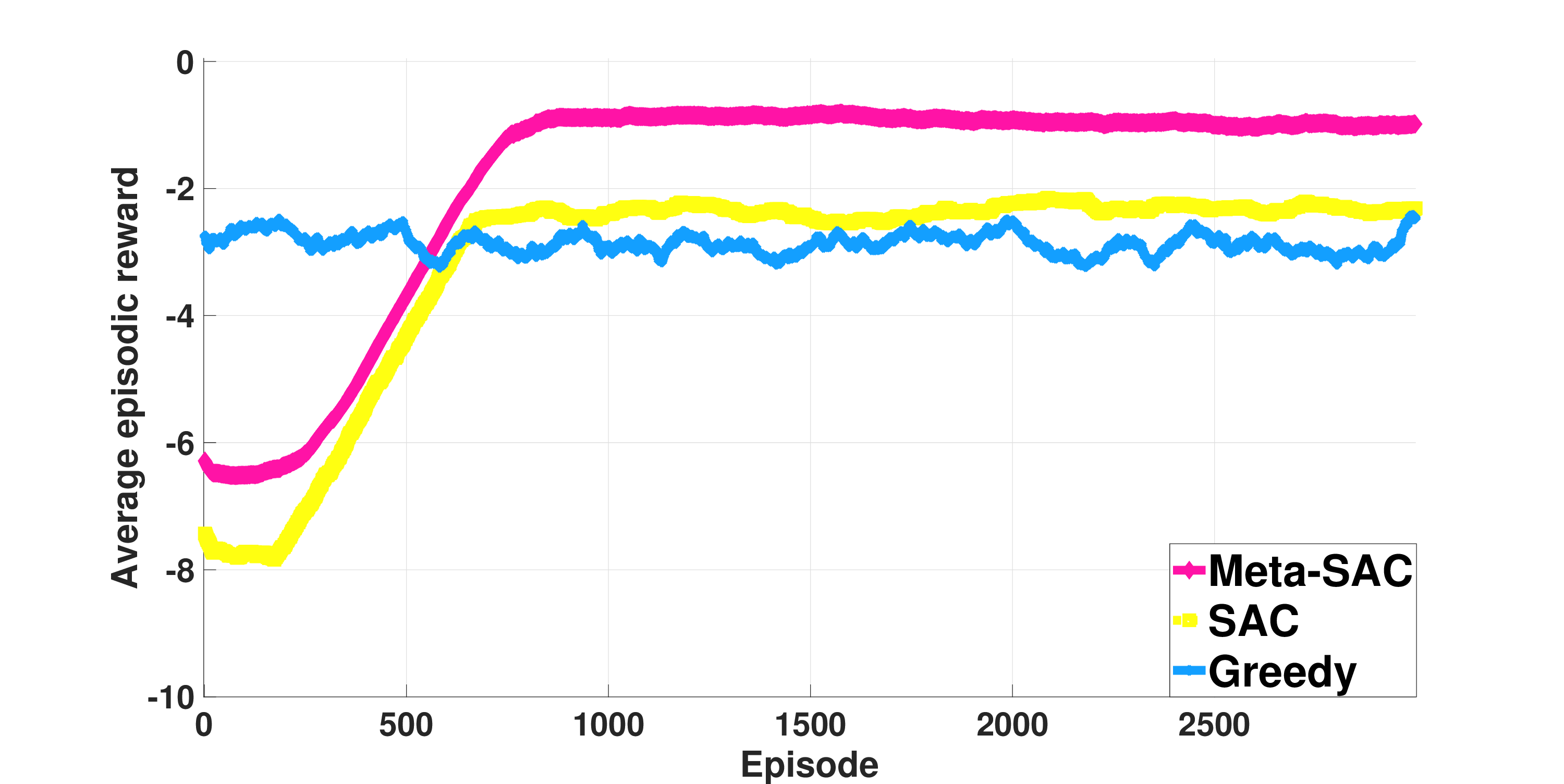}
 	\caption{Convergence behavior of the proposed resource allocation scheme.}
 	\label{fig:convergence} \vspace{-0.5cm}
\end{figure}
\begin{figure} 	
 	%\hspace{-0.5cm}
 	\includegraphics[scale=0.1550]{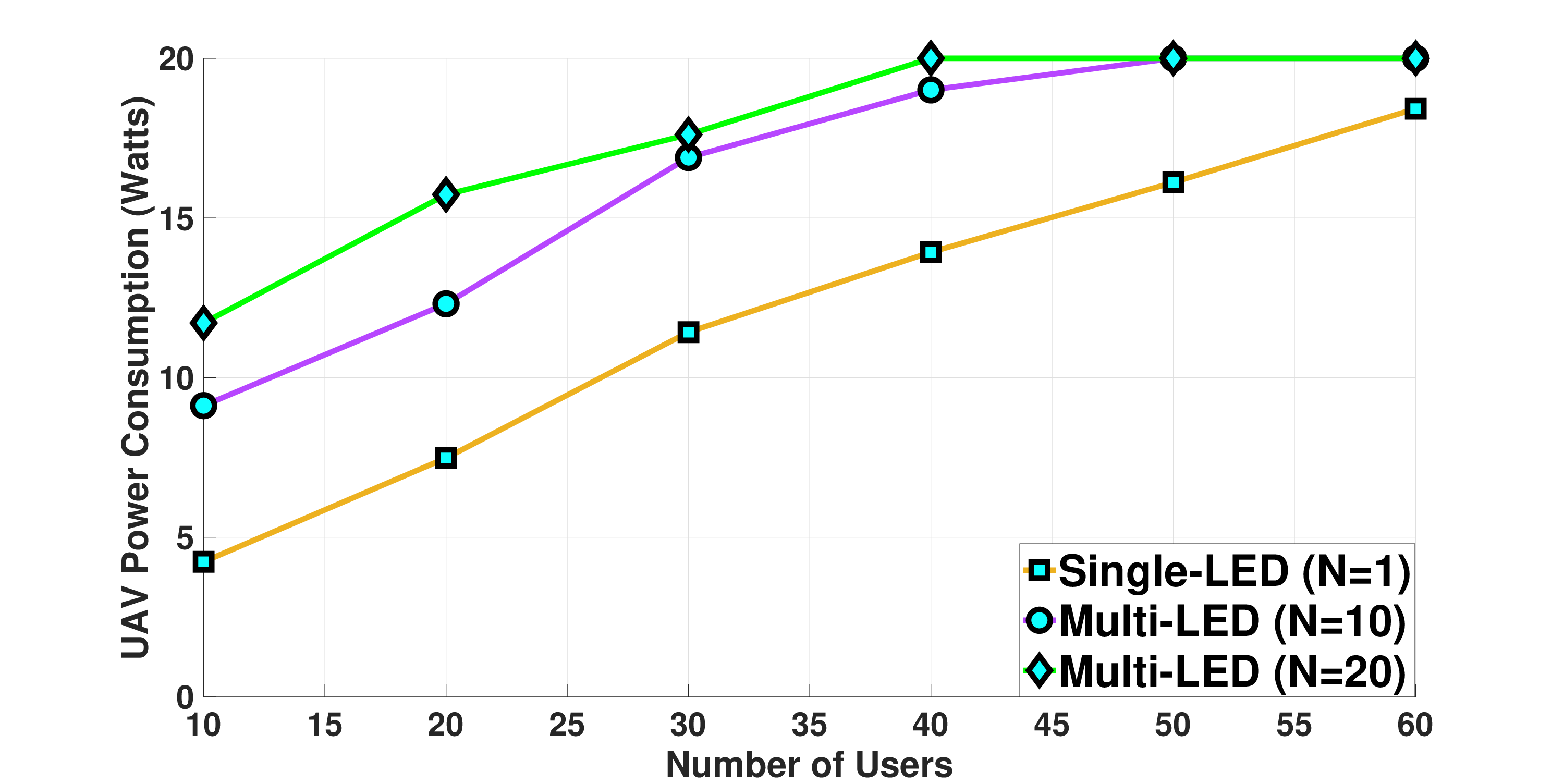}
 	\caption{Total power consumption of the UAV $P^{\text{Tot}}$ versus the number of users $K$.}
 	\label{fig:power} \vspace{-0.5cm}
\end{figure}
\begin{figure} 	
 	%\hspace{-0.5cm}
 	\includegraphics[scale=0.1550]{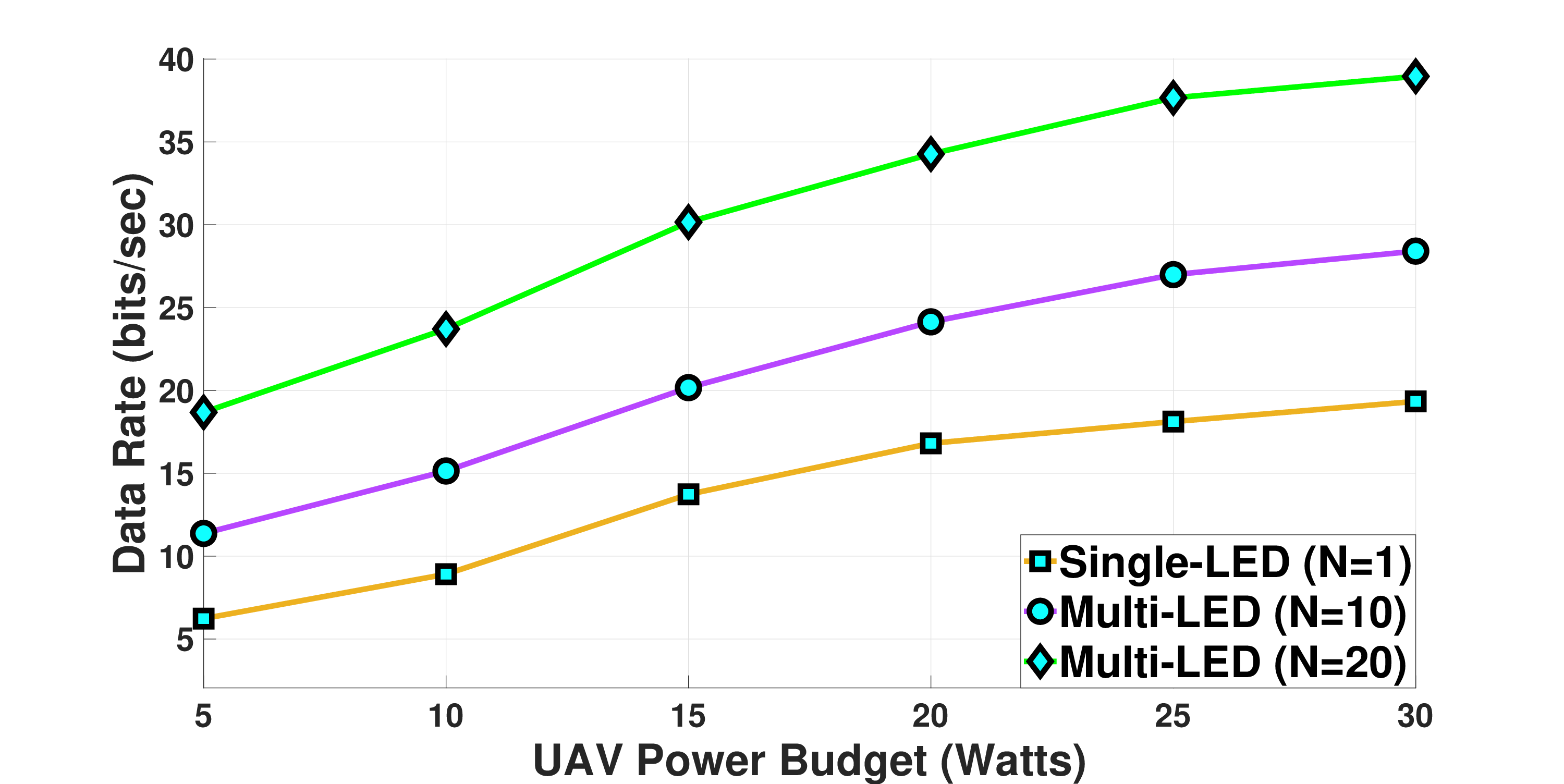}
 	\caption{Average achievable data rate versus power budget of the UAV $P_{\text{max}}$.}
 	\label{fig:data-rate} \vspace{-0.5cm}
\end{figure}
\section{Simulation Results} \label{Simulation}
This section evaluates the performance of the proposed UAV-MIMO-VLC system along with the corresponding resource allocation scheme. To that end, three baselines are considered, including a single-LED corresponded to \cite{data_rate}, as well as the baselines with $L$=10 and $L$=20 LEDs deployed at the UAV corresponding to our proposed system. The performance is assessed under three criteria, including power consumption of the UAV, as well as data rate and energy efficiency gained by the PD-equipped users. Simulations are performed with $K$=30 users, $R_{\text{min}}=2~\textrm{bits/s/Hz}$, $V_{\text{max}}=20~\text{m}/\text{s}$, $a_{\text{max}}=6~\text{m}/{\text{s}^2}$, $\textbf{q}_{\text{max}}=(150,150,100)~\text{m}$ and $\tau=1~\text{s}$. Other simulation parameters include $\Psi_{c}=60^{\circ}$, $n_{R}=1.5$, $\phi_{1/2}=60^{\circ}$, $\text{A}_\text{VLC}=1~\text{cm}^2$, $I_h=10~\text{mA}$, $I_l=0~\text{A}$, $I_0=5~\text{mA}$, $\ddot{\zeta}=1.2$, $\varphi=1$, $\zeta=1.225$, $N=10$,  $P_\text{max}=20~\text{W}$, $\rho=0.012$, $\delta=0.05$, $A_\text{U}=0.79$, $\Omega_\text{U}=400$, $R_\text{U}=0.05$, $\iota=1$, $W_\text{U}=100$, $v_{\text{UI}}=7.2$, $d_\text{U}=0.3$, $R_{\text{min}}=2~\textrm{bits/s/Hz}$, $V_{\text{max}}=10~\text{m}/\text{s}$, $a_{\text{max}}=6~\text{m}/{\text{s}^2}$, $\textbf{q}_{\text{max}}=(150,150,100)~\text{m}$, $\tau=1\text{s}$. 
\par Fig.~\ref{fig:convergence} shows the convergence behaviour of the proposed resource allocation scheme. Specifically, about 65\% lower power is consumed on average by the UAV while exploiting meta-learning. This significant reduction is attributed to the enhanced adaptability of the proposed meta-learning-enabled resource allocation scheme to frequent variations of the UAV-MIMO-VLC system. Moreover, this improvement is even more pronounced when compared to the greedy approach. 
%     \begin{table}[bp]
	%\vspace{-1 em}
%	\centering
%	\caption{Simulation Parameters}
%	\label{tab:my_label}
	%\vspace*{5pt}
%	\begin{tabular}{ l|l||l|l }
%		\textbf{Parameter} & \textbf{Value} & \textbf{Parameter} & \textbf{Value} \\
%		\hline
%		\hline		
%		\hline
%		$\Psi_{c}$   & $~60^{\circ}$ &
%		$n_{R}$ & $1.5$\\ 
%		\hline $ \phi_{1/2}$ & $~60^{\circ}$ &
%		$\text{A}_\text{VLC}$& $1~\text{cm}^2$ \%\ 
%		\hline $I_h$ & $10~\text{mA}$ &
%		$I_l$& $0~\text{A}$ \\
%		\hline $I_0$ & $5~\text{mA}$  &
%		$\ddot{\zeta}$ & $1.2$  \\
%		\hline $\varphi$ & $1$ &
%		$\zeta$ & $1.225$ \\
%		\hline $N$ & $10$  &
 %       $P_\text{max}$ & $20~\text{W}$ \\
  %      \hline $\rho$ & $0.012$  &
   %     $\delta$ & $0.05$ \\
    %    \hline $A_\text{U}$ & $0.79$  &
     %   $\Omega_\text{U}$ & $400$ \\
      %  \hline $R_\text{U}$ & $0.05$  &
       % $\iota$ & $1$ \\
       % \hline $W_\text{U}$ & $100$  &
        %$v_{\text{UI}}$ & $7.2$ \\
        %\hline $d_\text{U}$ & $0.3$  &
        %$R_{\text{min}}$ & %$2~\textrm{bits/s/Hz}$ \\
        % \hline $V_{\text{max}}$ & %$10~\text{m}/\text{s}$  &
        %$a_{\text{max}}$ & %$6~\text{m}/{\text{s}^2}$ \\
        %\hline $\textbf{q}_{\text{max}}$ & $(150,150,100)~\text{m}$  &
        %$\tau$ & $1\text{s}$ \\

        % $\hat{E}^{\text{gRA}}$ & 100 mJ &
        % $\hat{E}^{\text{SAT}}$ & 2000 mJ\\ \hline
        % $\mathcal{E}_i^{\text{gRA}}$ & 10 mJ &
        % $\mathcal{E}^{\text{SAT}}$ & 500 mJ\\ \hline
        % $\hat{Q}^{\text{gRA}}_i$ & 32\\ \hline

		% \hline $P^{\textrm{Har}}_\textrm{k,min}$ & $[10^{-8}-10^{-3}]~\text{w}$ &
		% $N_T$ & $6$ \\
	%	\hline
	%\end{tabular}
	%\label{tab2}
	%\vspace{-0.75 em}
%\end{table}
\par Fig.~\ref{fig:power} shows the effect of increasing the number of PD-equipped users on power consumption of the UAV. Evidently, by increasing the number of users, fulfilling their data rate requirement, as dictated in C$_1$, induces more power consumption to the UAV. Quantitatively, a UAV with a LED array with $N$=10 LEDs consumes 40\% more power in average, compared to a UAV employing a single-LED, when $K$=10. However, by increasing the number of users, the gap between the baselines diminishes as all baselines gradually approach the power budget of the UAV. For instance, when $K$=30, the gap between the aforementioned baselines falls under 8\%. This observation underscores the effectiveness of the proposed UAV-MIMO-VLC system in dense networks. Another insight, drawn from Fig.~\ref{fig:power} corresponds to the limited gap between the baselines with $N$=10 and $N$=20. This observation highlights the influence of the adopted hybrid dimming mechanism, which effectively controls the power consumption and promises the deployment of massive LED arrays to gain higher data rates.  
\begin{figure} 	
 	%\hspace{-0.5cm}
 	\includegraphics[scale=0.1550]{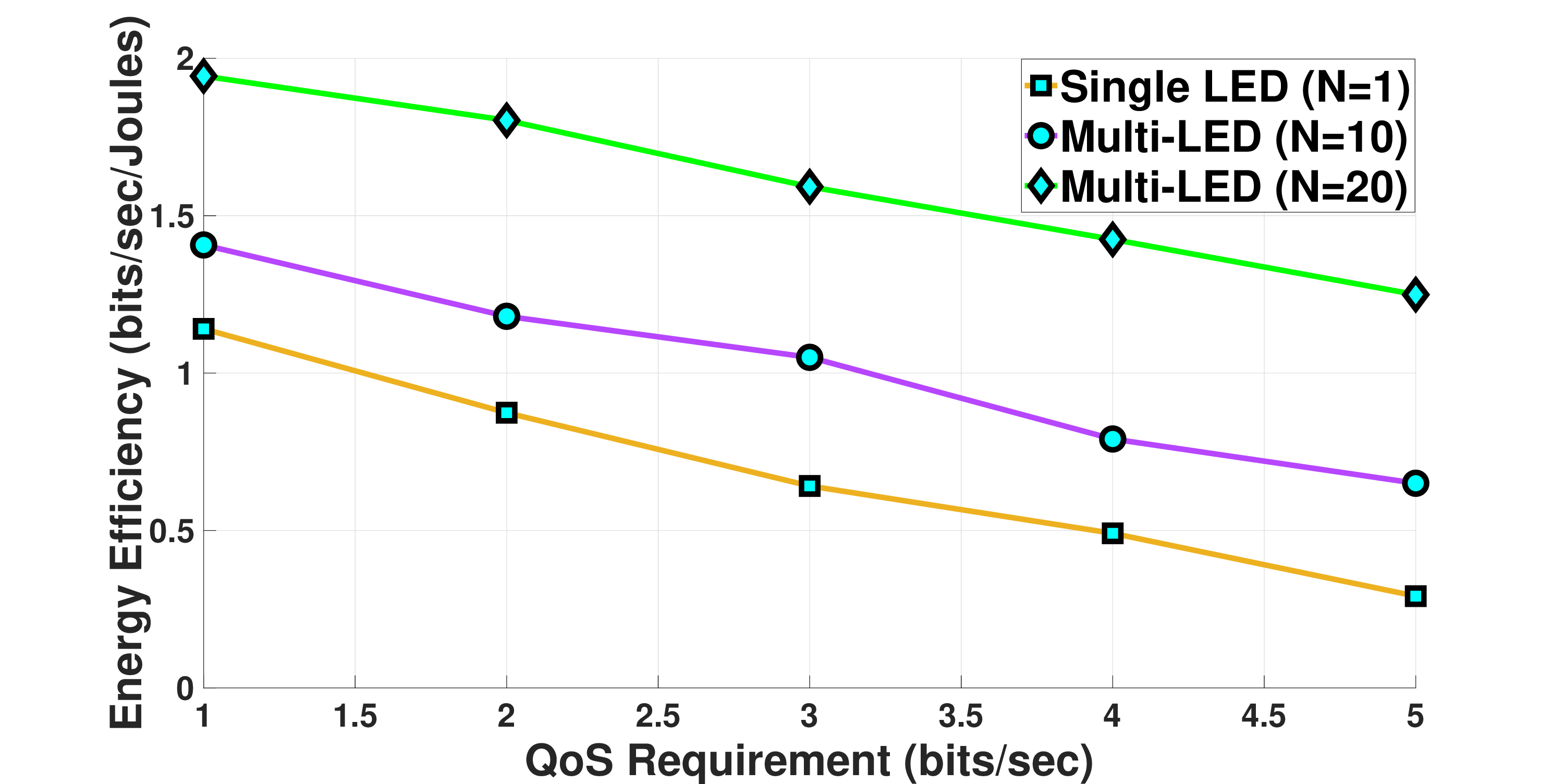}
 	\caption{Average achievable energy efficiency versus $R_{\text{min}}$.}
 	\label{fig:energy-efficiency} \vspace{-0.5cm}
\end{figure}
\par Fig.~\ref{fig:data-rate} depicts the impact of the power budget of the UAV $P_{\text{max}}$, on maximizing the achievable data rate of the UAV-MIMO-VLC system. The proposed resource allocation scheme effectively exploits the available power budget, such that the inter-user interference in \eqref{rate} is managed well and the overall achieved data rate of the network increases. More importantly, the single-LED baseline obtains much lower data rate, compared to the baselines with $N$=10 and $N$=20 LEDs. On average, around 47\% gain is achieved, when comparing the $N$=10 and $N$=1 scenarios. This observation underscores the cost-effectiveness of the deployment of LED array at the UAV in comparison with a single-LED from the network data rate standpoint. Moreover, the gap between the baselines $N$=10 and $N$=20 is notably larger than that between $N$=1 and $N$=10. This suggests that deploying massive LED arrays can better exploit the abundant capacity of the VLC band to fulfill the explosive data rate requirements of upcoming wireless networks. 
%As dictated by C$_2$, relaxing a constraint expands the solution space of the optimization problem $\mathcal{P}_1$, as the result of which, the overall data rate of the network as the objective improves. 
\par Fig.~\ref{fig:energy-efficiency} shows the performance of the UAV-MIMO-VLC system, aimed at maximizing the overall energy efficiency, defined as a trade-off between the achievable data rate and power consumption of the system. In essence, assurance of more QoS requirement for users induces a user fairness within the network, thereby degrades the overall achievable data rate of the network. Conversely, the UAV has to spend more power to meet higher QoS requirements. Under this argument, the more QoS requirement for users is guaranteed, the lower energy efficiency is achieved. Deploying a LED array with a greater number of LEDs at the UAV achieves much more energy efficiency, which underscores the effectiveness of the proposed UAV-MIMO-VLC system. For instance, in average, 34\% better energy efficiency is achieved by the baseline with $N$=10, compared to the single-LED baseline $N$=1.
\par In table I, we have investigated the energy efficiency of the system, as defined in \eqref{EE}, by increasing the number of LEDs from 1 to 10 and 20. The comparison baselines include Meta-SAC, SAC, PPO \cite{HD-1} and random. Obviously, the proposed Meta-SAC outperform others, thanks to the generalization power, meta-learning contributes to this scheme. Furthermore, compared to random scheme, lacking the optimization of resources, other baselines exhibit a superior performance. \vspace{-0.2cm}
\section{Conclusions} \label{Conclusions}
This paper proposes a UAV-MIMO-VLC system together with an adaptive resource allocation framework. The numerical results indicates that equipping the UAV with a LED array rather than a single-LED, significantly improves the data rate and energy efficiency of the network, yet at the cost of increased power consumption.
\begin{table}[h]
\centering
\captionsetup{font=small}
\captionsetup{justification=centering}
	\caption{Average achievable energy efficiency (bps/Watts)}
	\begin{center}
		\begin{tabular}{|l|l|l|l|l|l|}
			\hline
		\diagbox[width=14em]{Scheme}{\#LEDs} & {1} & {10} & {20}\\ 
			\hline\hline
			Meta-SAC & 2.37 & 3.41 & 4.93\\
			\hline
			SAC & 1.85 & 2.62 & 3.82\\
			\hline
   PPO \cite{HD-1} & 1.34 & 2.11 & 2.94\\
   \hline
   Random & 0.46 & 1.13 & 1.91\\
			\hline
		\end{tabular}
		\label{table:comp1}
	\end{center}\vspace{-0.4cm}
\end{table}\vspace{-0.1cm}
\section{Acknowledgment}
Sinem Coleri acknowledges the support of Ford Otosan. The work of Eduard A. Jorswieck was supported by the Federal Ministry of Education and Research of Germany in the Program of “Souveraen. Digital. Vernetzt.” Joint Project 6G-Research and Innovation Cluster (6G-RIC) under Project 16KISK031.

\bibliographystyle{ieeetr} % Choose a style: plain, alpha, unsrt, etc.
% \bibliography{main.bbl} % Name of your .bib file (without .bib extension)
\vspace{-0.3cm}
\balance
		
\end{document}